\documentclass{ws-p8-50x6-00}
\begin{document}

\title{Effects of Strong Magnetic Fields in Neutron Star and Gravitational Wave Emission}

\author{G. F. Marranghello}

\address{Physics Institute, Universidade Federal do Rio Grande do Sul, Porto Alegre, Brazil
\\ gfm@if.ufrgs.br}


\maketitle

\abstracts{I review in this work the properties of nuclear and
    quark matter equation of state under the influence of a strong magnetic field. We analyse
    the results when they are applied to compact stars description. Finally, we
    describe the emission of gravitational wave from a compact star
    that goes over a phase transition.}

\section{Introduction}

The determination of the nuclear matter equation of state and the
appearence of new phases at high densities is one of the main
goals in present nuclear physics. Nowadays, huge recently planned
experiments have brought excellent expectations on these
determinations. RHIC and LHC are the most promissing tools for direct
observations of the high density matter behaviour, from heavy ion
collisions. Another important experiments are linked to the
observation of neutron star properties, once they are believed to
behave like giant hypernucleus and/or giant quark bags. In this direction, 
Chandra X-Ray Observatory and BeppoSax have done an
excellent job analysing the data originated by X-Ray or thermal
emission from pulsars. We emphasize the data that indicates
important results about more massive neutron stars (M$\sim$2.0-2.2M$_\odot$) 
and others about smaller stars with radius $\sim 30\%$ smaller then ordinary 
neutron star radius. New astrophysical tools are related to the emission
of gravitational waves from these extremely dense objects and
shall be observed by the american LIGO or by the french-italian
collaboration VIRGO.

Such a variety of experiments being located for the seek of
nuclear matter equation of state and the quark-gluon plasma phase
transition implies in a large ammount of theoretical works which
are capable to improve the experiments capabilities, direct them
to better targets and have their payback with results that may
prove theories. Nowadays one can find different works described in
many papers, not only because of theoretical physicist's
criativity but also because of the many different approachs high
energy physics support. One can go from the complex analytical
quantum chromodynamics (QCD), step on the lattice simulations,
drive through feynman diagrams, and get wherever he wants. They
all are validy and the scope of this work do not include a
discussion of the best way of doing nuclear physics or the best
model to fit the bulk properties of matter. We intend to use a
quantum hadrodynamics (QHD)\cite{Serot1} based model and the MIT\cite{MIT} 
bag model to describe compact stars main properties and trace a consistent 
path from bulk nuclear matter description to neutron star properties and the
gravitational wave emission.

\section{Nuclear Matter Equation of State\label{eos}}

One consistent, well based and effective model to describe nuclear
matter is based on the pioneered work by Walecka which consists
one of the basis of nuclear matter effective field theory. Further
developments were done by many authors from whose we extract the
works by Boguta and Bodmer\cite{Boguta}, Zimanyi and Moszkowski\cite{Zimanyi} and Taurines
et. al.\cite{Taurines}, developed in the following years.

We have used the model as described by the lagrangian density below 

\begin{eqnarray}
{\cal L}    &=& \sum\limits_{B}   \bar{\psi}_{B} [( i\gamma_\mu (\partial^\mu- g_{\omega B} \omega^{\mu}) -
(M_B-g_{\sigma B} \sigma)
]\psi_B \nonumber \\ && - \sum\limits_{B}   \bar{\psi}_{B} [
\frac12 g_{\varrho B} \mbox{\boldmath$\tau$} \cdot \mbox{\boldmath$\varrho$}^\mu] \psi_B
+\frac{bM}{3}\sigma^3+\frac{c}{4}\sigma^4 \nonumber \\ &&+\frac12(\partial_\mu \sigma \partial^\mu \sigma   -
{m_\sigma^2} \sigma^2)  - \frac14  \omega_{\mu \nu}  \omega^{\mu \nu}   + \frac12 {m_\omega^2}
 \omega_\mu \omega^\mu  \nonumber \\
&& -   \frac14 \mbox{\boldmath$\varrho$}_{\mu \nu} \cdot \mbox{\boldmath$\varrho$}^{\mu \nu} +  \frac12m_\varrho^2
\mbox{\boldmath$\varrho$}_\mu
 \cdot  \mbox{\boldmath$\varrho$}^\mu \nonumber \\
&&+\sum\limits_{l}   \bar{\psi}_{l} [i \gamma_\mu \partial^\mu   - M_l] \psi_l \,\, .
\end{eqnarray}
which includes the baryonic octet (p, n, $\Lambda$,$\Sigma^+$, $\Sigma^0$, $\Sigma^-$, $\Xi^-$,
$\Xi^0$) coupled to mesons ($\sigma, \omega, \varrho$), and free leptons (e,
$\mu$). The scalar and vector coupling constants, g$_{\sigma}$, g$_{\omega}$ and
the coeficients b, c are determined through the nuclear matter bulk properties
fitting, i. e., binding energy E$_b$ (= -16.3 MeV), compression modulus K (= 240
MeV) and effective nucleon mass $M^* = M-g_{\sigma} \bar{\sigma}$ (= 732
MeV) at saturation density $\rho_0$ (=0.153
fm$^{-3}$). The g$_{\varrho}$ coupling constant is determined through the
nuclear matter asymmetry coefficient, a$_4$ (= 32.5 MeV). All the coefficient
values are described in ref.\cite{Gfm,Gfm2}.

\begin{center}
\begin{figure}[htb]
\vspace*{10pt} 
\vspace*{1.4truein}             
\vspace*{10pt} \parbox[h]{4.5cm}{ \includegraphics{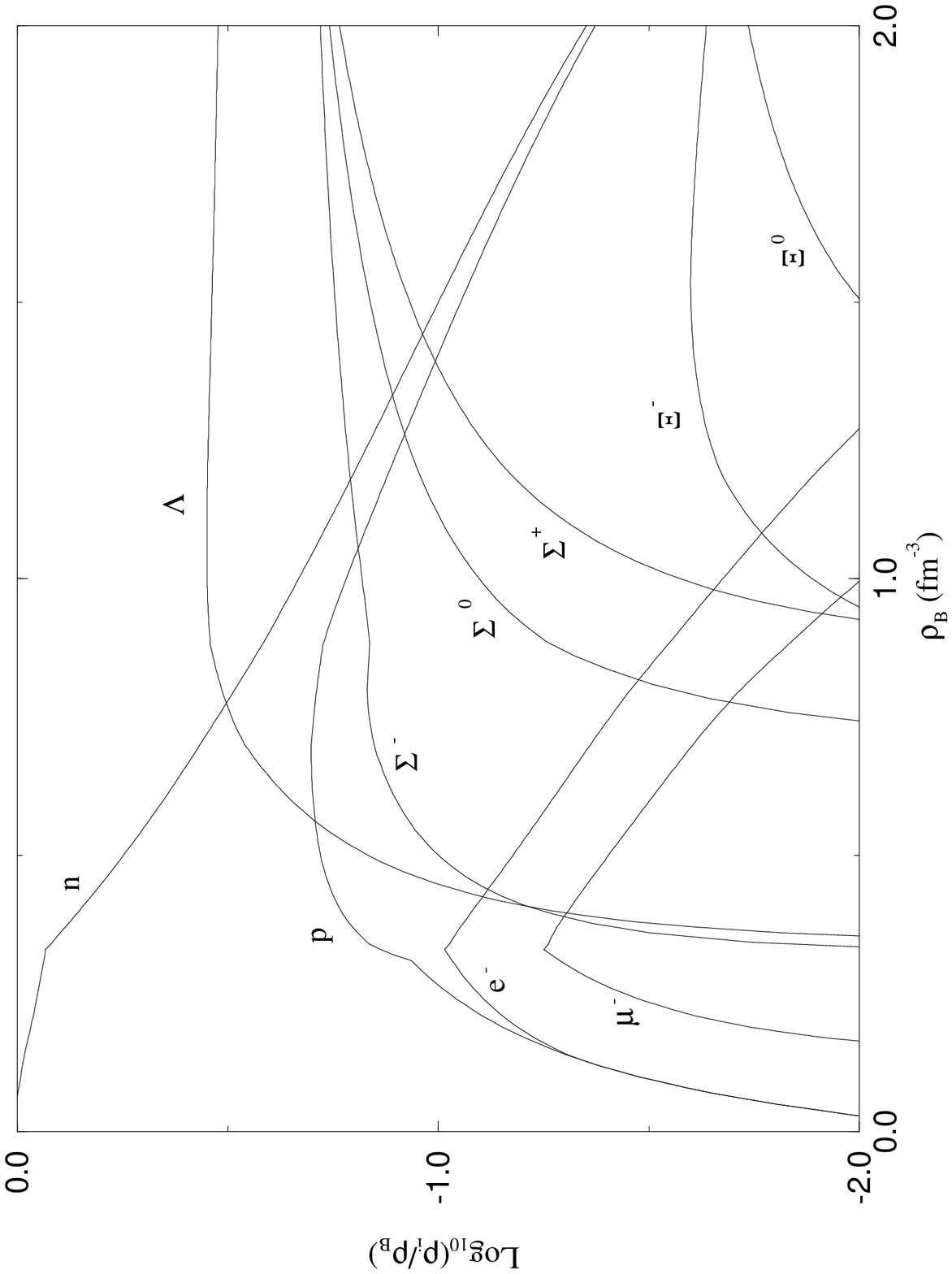}}
\vspace{-10pt} \caption{Particle distribution as function of baryon density.
\label{pd}}
\end{figure}
\end{center}

Applying standard technics of field theory and solving the equations
at a mean-field level one can extract expressions for the
thermodynamical quantities of pressure, $p$, energy density,
$\epsilon$ etc. and verify the suitability of this model results
to the experimental values of compression modulus, $K$ and
effective nucleon mass, $M^*$, resulting in excellent results for
neutron star masses and radii, which can be compared to previous works\cite{Glendenning2,Gfm}.
In figure \ref{pd} we describe the behaviour of the density of each different
particle described by the model as a function of the total baryon density.

\section{Phase Transition to Quark Matter\label{qm}}

The nuclear to quark matter phase transition is expect to occur at high
densities and/or high temperatures, restoring the chiral symmetry. We follow in
this section the work by Heiselberg and Hjorth-Jensen\cite{Heiselberg} whose
perfectly described the conditions for the occurence, or not, of a mixed phase
composed by hadrons and deconfined quarks simmultaneously. First claimed by
Glendenning\cite{Glendenning}, the occurence of a mixed phase where quark structures are immersed
into a hadron matter (or vice-versa at higher densities), is now uncertain. This
work will not clarify this topic, once it is model-dependent, however, as an
application of this study we intend to propose the possible difference on the
detectation of such phases inside hybrid stars.

Generally, the properties of the transitions with one or more than
one conserved charges are quite different. The most important
feature is that pressure may be constant or vary contineously with
the proportion of phases in equilibrium. Reviewing Heiselbergs
work, one can search the answer of which kind of transition is
more energetically favored through an analysis of the bulk energy
gained in such a transition and the Coulomb and the surface energy
relation, as we stress below.

Assuming the model described in the previous section and the MIT bag model for
the quark matter, 

\begin{eqnarray}
\Omega & = & \sum_{q=u,d,s}\frac{-1}{4\pi^2}\left[\mu_q {k_F}_q(\mu_q^2-\frac52
m_q^2)+\frac32 m_q^4 ln\left(\frac{\mu_q+{k_F}_q}{m_q}\right)\right] \nonumber
\\ &+& \frac{2\alpha_c}{4\pi^3}\left[3\left[\mu_q {k_F}_q-m_q^2-m_q^2
ln\left(\frac{\mu_q+{k_F}_q}{m_q}\right)\right]^2 
-2{k_F}_q^4 - 3m_q^4 ln^2\left(\frac{m_q}{\mu_q}\right)\right] \nonumber \\
&\times&\left[6ln\left(\frac{\rho_r}{\mu_q}\right) \left(\mu_q {k_F}_q m_q^2
-m_q^4 ln\left(\frac{\mu_q+{k_F}_q}{m_q}\right)\right)\right]+B,
\end{eqnarray}
where $\Omega$ is the thermodynamical potential for the MIT bag model,
we relate the energy densities in order to calculate the bulk
energy. Giving the large uncertainties in the estimates of bulk and surface
properties, one cannot claim that droplet phase is favored or not, once it
depends crucially on nuclear and quark matter properties.

If droplet sizes and separations are small compared with Debye screening
lenghts, the electron density will be uniform to good approximation, and the
Debye screening lenght $\lambda_D$ is defined by

\begin{equation}
\frac{1}{\lambda_D^2}=4\pi\sum_i Q_i^2\left(\frac{\partial
n_i}{\partial\mu_i}\right)_{n_j,j\ne i}
\end{equation}
where $n_i$, $\mu_i$ and $Q_i$ are the number density, chemical potential and
charge of particle species $i$.

In the system we consider, screening effects can be estimated and, if the
characteristic spatial scales of structures are less than about 10 fm for the
nuclear phase, and less than about 5 fm for the quark phase, screening effects
will be unimportant, and the electron density will be essentially uniform. In
the opposite, the total charge densities in bulk nuclear and quark matter will
both vanish.

When quark matter occupies only a small fraction, $f=\frac{V_{QM}}{V_{QM}+V_{NM}}$, of
the total volume, quarks will form spherical droplets. The surface energy per
droplet is given by $\epsilon_S=\sigma 4\pi R^2$, where $\sigma$ is the surface
tension, and the Coulomb energy is

\begin{equation}
\epsilon_C=\frac{16\pi^2}{15}(\rho_{QM}-\rho_{NM})^2 R^5.
\end{equation}
Minimizing the energy density with respect to $R$ one obtain the usual result
$\epsilon_s=2\epsilon_C$ and find a droplet radius

\begin{equation}
R=\left(\frac{15}{8\pi}\frac{\sigma}{(\rho_{QM}-\rho_{NM})^2}\right)^2.
\end{equation}

Our results, considering the models described before, cannot be definitive
before a high precise determination of the surface tension, $\sigma$. However,
for $\sigma\sim 70-100 MeV$, a mixed phase is unfavored and the star will have a
density discontinuty. The result is a larger radius to support the mixed phase
and the equation of state for the constant pressure phase transition is
presented below.

\begin{center}
\begin{figure}[htb]
\vspace*{10pt} 
\vspace*{1.4truein}             
\vspace*{10pt} \parbox[h]{4.5cm}{ \includegraphics{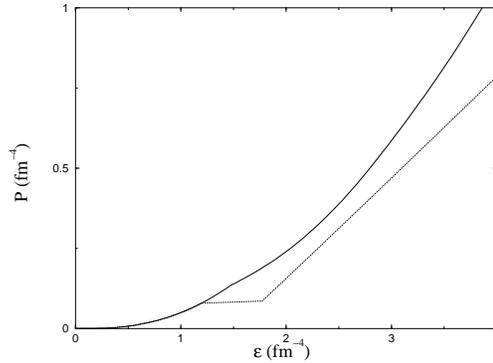}}
\vspace{-10pt} \caption{Equation of state for pure neutron star matter (solid
  line) and for the transition to quark matter (dotted line).
\label{}}
\end{figure}
\end{center}

\section{Magnetic Field Effects\label{b}}
Several indenpendent arguments link the class of {\it Soft
$\gamma$-Ray Repeaters} (SGRs) and perhaps anomalous X-ray pulsars
with neutron stars having ultra strong magnetic fields. In
addition, two of four known SGRs directly imply, from their
periods and spin-down rates, surface fields in the range
$2-8\times 10^{14}G$. The magnitude of the magnetic field strength
needed to dramatically affect neutron star structure and an
estimate done by Lai and Shapiro\cite{Lai} yields $B\sim 2\times
10^{18}\frac{M/1.4M_\odot}{(R/10km)^2}$ in the interior of neutron
stars.

In this section we investigate the effects of very strong magnetic
fields upon the equation of state of dense matter in which
hyperons and quarks are present. In the presence of a magnetic
field, the equation of state above nuclear saturation density is
significantly affected both by Landau quantization and magnetic
moment interactions, but only for fields strengths $B>10^{18}G$.

The previously presented lagrangian density in section \ref{eos} can be
re-written in order to include the magnetic field as

\begin{eqnarray}
 {\cal{L}} & = &
 \sum\limits_{B} \bar{\psi}_B [i \gamma_\mu \partial^\mu - (M_B-
g_{\sigma B}^\star \sigma)-g_{\omega  B}^\star \gamma_\mu \omega^\mu
 +q_b\gamma_\mu A^\mu \nonumber \\ & - &\kappa_b\sigma_{\mu\nu}F^{\mu\nu}]\psi_B   -  \sum\limits_{B}
\psi_B [\frac12 g_{\varrho B}^\star \gamma_\mu \mbox{\boldmath$\tau$}\cdot
 \mbox{\boldmath$\varrho$}^\mu] \psi_B \nonumber \\
& + & \sum\limits_\lambda \bar{\psi}_\lambda [i \gamma_\mu \partial^\mu -
m_\lambda +q_l\gamma_\mu A^\mu] \psi_\lambda \nonumber \\ & + & \frac12(\partial_\mu \sigma \partial^\mu \sigma - {m_\sigma}^2 \sigma^2)
- \frac14 \Omega_{\mu \nu} \Omega^{\mu \nu} - \frac14 F_{\mu \nu} F^{\mu \nu}
 \nonumber \\ & + & \frac12 {m_\omega}^2 \omega_\mu \omega^\mu \nonumber -  \frac14
\mbox{\boldmath$\varrho$}_{\mu \nu} \cdot \mbox{\boldmath$\varrho$}^{\mu \nu} + \frac12m_\varrho^2
\mbox{\boldmath$\varrho$}_\mu \cdot \mbox{\boldmath$\varrho$}^\mu.
\end{eqnarray}

For detailed and well founded texts on magnetic field effects the
authors recommend ref.\cite{Broderick,Broderick2}, from where one
can find the energy spectrum for the protons given by

\begin{eqnarray}
E_p & = &\sqrt{k_z^2+\left(\sqrt{{M_p^*}^2+2*(n+\frac12+\frac{s}{2})q_pB}+s\kappa_pB\right)^2}\nonumber
\\ & + &g_\omega\omega_0-\frac12g_\varrho{\bf\varrho}_0.
\end{eqnarray}

In a strong magnetic field, particles motion is perpendicular to the
field lines and are quantized to give discrete {\it Landau orbitals},
and the particles behave like a one-dimensional gas rather than a
three-dimensional gas and, the higher is the field, the lower is
the number of occupied {\it Landau levels} The energy of a charged
particle changes significantly in the quantum limit if the
magnetic field $H\ge H_c$. For electrons, $H_c\sim 4\times 10^{13}
G$, $u$ and $d$ (massless quarks), $H_c \sim 4\times 10^{15} G$,
$s$-quark, $H_c\sim 10^{18} G$ and protons, $10^{20} G$.

Considering the phase transition aspects, a strong magnetic field unfavor the
mixed hadron-quark phase once it increases the eletron number in the nuclear
phase leaving the quark phase praticaly unchanged. This increases the
droplet size as it was discussed in the previous session and in ref.\cite{Bandy}.

\section{Compact Stars\label{cs}}

The first step through an analysis of compact stars recall the simpler solution
to the general relativistic Einsteins equations which represents the static and
spherical symmetric stars, the Schwarschild solution, written as the
Tolman-Oppenheimer-Volkoff equations\cite{Tolman,Oppenheimer}.
The TOV equations describe the structure of a static, spherical and isotropic
star wit the pressure $p(r)$ and the
energy density $\epsilon(r)$ reflecting the underlying nuclear model. The TOV
equations involve various constraints and boundary
conditions: they must be evaluated for the
initial condition $\epsilon(0)=\epsilon_c$ (with $\epsilon_c$ denoting the
central density) and $M(0)=0$ at $r=0$;
the radius R of the star is determined under the condition that on its surface
the pressure vanishes, $p(r)|_{r = R}= 0$.

The condition for chemical equilibrium for neutron stars are:
\begin{equation}
\mu_i = b_i\mu_n-q_i(\mu_{\ell})
\end{equation}
where $\mu_i$ and $\mu_{\ell}$ stand for the baryon and lepton
chemical potentials, respectively;  $b_i$ is the baryon number and
the baryon and lepton electrical charges are represented by $q_i$.

The corresponding equations for baryon number and electric charge conservation are:
\begin{equation}
\rho_{baryonic} = \sum_B \frac{k_{F,B}^3}{3 \pi^2} \, ,
\end{equation}
and
\begin{equation}
\sum_B q_{e,B} \frac{k_{F,B}^3}{3 \pi^2} - \sum_{\ell} \frac{k_{F,\ell}^3}{3 \pi^2}=0 \, .
\end{equation}

Here one can visualize the importance and the effects of quark and
nuclear matter parameters determination, once they will have
reflects on the properties of compact stars. First, quark stars
maximum mass and radius are directly governed by the value of the
bag constant as:

\begin{equation}
M=\frac{1.964 M_\odot}{\sqrt{B_{60}}} \, \, \, R=\frac{10.71 km
}{\sqrt{B_{60}}},
\end{equation}
where $B_{60}=B/(60 MeV/fm^3)$ in the massless quarks case. In the same sense, the strong
coupling constant $\alpha_c$ is related to the quark star
properties. Both $B$ and $\alpha_c$ will determine whether or not
not the hadron-quark phase transition will take place. The sequence of neutron
stars (mass-energy density relation) and the energy density in the
interior of a neutron star is presented in the figures below.

\begin{center}
\begin{figure}[htb]
\vspace*{10pt} 
\vspace*{1.4truein}             
\vspace*{10pt} \parbox[h]{4.5cm}{ \includegraphics{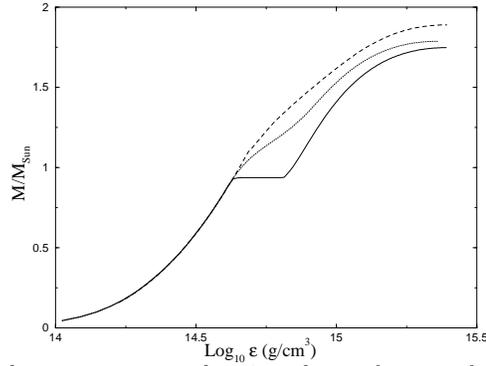}}
\vspace{-10pt} \caption{Masses of neutron star as a function of central energy
  density. The hybrid star without mixed phase (solid line), with mixed phase
  (dotted line) and the pure neutron star (dashed line) are presented.
\label{}}
\end{figure}
\end{center}

\begin{center}
\begin{figure}[htb]
\vspace*{10pt} 
\vspace*{1.4truein}             
\vspace*{10pt} \parbox[h]{4.5cm}{ \includegraphics{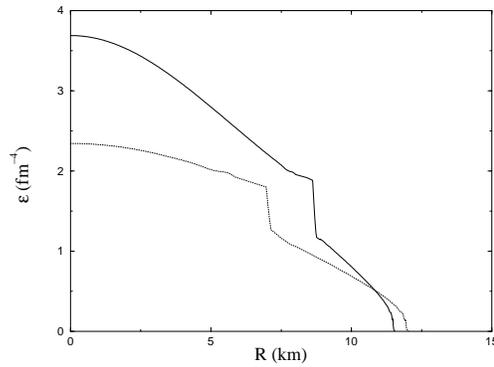}}
\vspace{-10pt} \caption{Energy density distribution inside the neutron star for
  two different central energies.
\label{}}
\end{figure}
\end{center}

The nuclear matter coupling constants $g_{B\sigma}$, $g_{B\omega}$
and $g_{B\varrho}$ present the same sensibility and uncertainties.
Nuclear matter results for effective mass compression modulus and
particle distribution play also an important role on the phase
transition and star properties and, consequently, on the emission
of gravitational wave as we will discuss in the following
sections.

The magnetic field effects are presented in the following figures of equation of
state, neutron star masses and particle distribution.

\begin{center}
\begin{figure}[htb]
\vspace*{10pt} 
\vspace*{1.4truein}             
\vspace*{10pt} \parbox[h]{4.5cm}{ \includegraphics{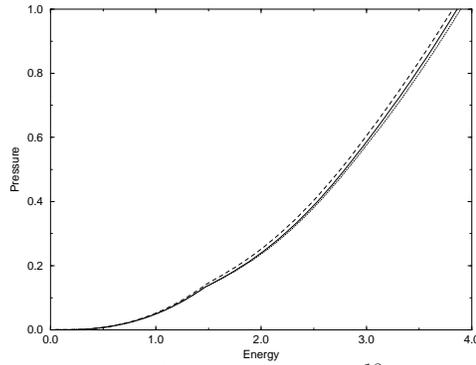}}
\vspace{-10pt} \caption{Equation of state at B=0 (solid line) and B=10$^{18}$G
  with (dotted line) and without (dashed line) the effects of anomalous magnetic moment.
\label{}}
\end{figure}
\end{center}

\begin{center}
\begin{figure}[htb]
\vspace*{10pt} 
\vspace*{1.4truein}             
\vspace*{10pt} \parbox[h]{4.5cm}{ \includegraphics{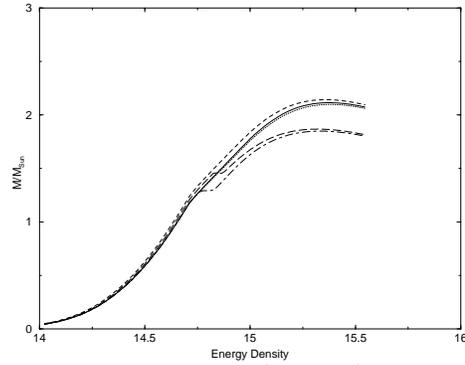}}
\vspace{-10pt} \caption{Masses of pure neutron star at B=0 (solid line) and
  B=10$^{18}$G with (dotted line) and without (dashed line) the effects of
  anomalous magnetic moment and the masses of hybrid stars at B=10$^{18}$G (long
  dashed line) and B=$5\times 10^{18}$G (dot-dashed line) as functions of central energy density.
\label{}}
\end{figure}
\end{center}

\begin{center}
\begin{figure}[htb]
\vspace*{10pt} 
\vspace*{1.4truein}             
\vspace*{10pt} \parbox[h]{4.5cm}{ \includegraphics{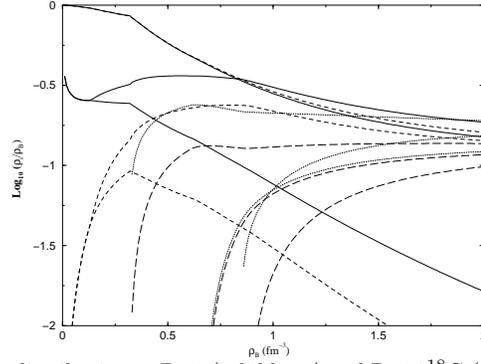}}
\vspace{-10pt} \caption{Particles distribution at B=0 (solid lines) and
  B=10$^{18}$G (dotted line) as functions of the baryonic density.}
\end{figure}
\end{center}

\section{The Conversion of Neutron Stars to Hybrid Stars\label{nstoss}}

The transition from configuration H to HQ may occur through the formation of
a mestastable core, built up by an increasing central density. The increase
in the central density may be a consequence of
a continuous spin-down or other different mechanisms the star could suffer. This transition releases energy, exciting mainly the radial modes of
the star. These modes do not emit GWs, unless when coupled with
rotation, a situation which will be assumed here.

Strange matter is assumed to be absolutely stable and a seed of
strange matter in a neutron star would convert the star into a
hybrid or strange star. The speed at which this conversion occurs
was calculated by Olinto \cite{Olinto}, taking into account the
rate at which the down- and strange-quark Fermi seas equilibrate
via weak interactions and the diffusion of strange-quarks towards
the conversion front.

Accordingly to \cite{Olinto}, cold neutron stars can convert to
strange star matter with speeds ranging from $5 km/s$ to $2\times
10^4 km/s$. The outcome of such an event would emit an incredible
amount of energy in $0.5 ms$ to $2 s$. We apply this discussion to
the partial conversion of a neutron star to strange matter,
forming a hybrid star.

However, the emission of a gravitational wave, which will be
discussed in the next section, can only {\it recognize} the phase
transition after a structural rearrangement of the star, which
shall reduce its radius and gravitational mass, conserving its
total baryon number. Such mechanism has already been studied by
Bombaci \cite{Bombaci} for the emission of $\gamma$-ray bursts.

\section{Gravitational Wave Emission\label{gw}}

In order to simplify our analysis, we will consider that most of the mechanical energy
is in the fundamental model. In this case, the gravitational strain amplitude can be
written as
\begin{equation}
h(t) = h_0e^{-(t/{\tau_{gw}}-\imath\omega_{0} t)}
\end{equation}
where h$_0$ is the initial amplitude, $\omega_0$ is the angular frequency of the mode and $\tau_{gw}$ is the
corresponding damping timescale. The initial amplitude is related to the total energy E$_g$ dissipated under the
form of GWs by the relation\cite{pacheco01}
\begin{equation}
h_0 = {{4}\over{\omega_0 r}} {\left[{GE_g}\over{\tau_{gw} c^3} \right]^{1/2}}
\end{equation}
where G is the gravitational constant, c is the velocity of light and r is the distance to
the source.

Relativistic calculations of radial oscillations of neutron star
with a quark core were recently performed by Sahu et
al.\cite{sahu01}. However, the relativistic models computed by
those authors do not have a surface of discontinuity  where an
energy jump occurs. Instead a mixing region was considered, where
the charges (electric and baryonic) are conserved globally but not
locally\cite{Glendenning}. Oscillations of star models including
an abrupt transition between the mantle and the core were
considered by Haensel\cite{haensel89}, Miniutti\cite{miniutti} and Marranghello\cite{Gfm2}.
However a Newtonian treatment was adopted and the equation of
state used in the calculations does not correspond to any specific
nuclear interaction model. In spite of these simplifications,
these hybrid models suggest that {\it rapid} phase transitions, as
that resulting from the formation of a pion condensate, proceed at
the rate of strong interactions and affect substantially the mode
frequencies. However, the situation  is quite different for {\it
slow} phase transitions (the present case), where the mode
frequencies are quite similar to those of a {\it one-phase}
star\cite{haensel89}. In this case, scaling the results of
\cite{haensel89}, the frequency of the fundamental mode
(uncorrected for gravitational redshift) is given approximately by
\begin{equation}
\nu_0 \approx 63.8 \left[ {{(M/M_{\odot})}\over{R^3}} \right]^{1/2} \,\ \,\ kHz
\end{equation}
where the mass is given in solar units and the radius in km.

Once the transition to quark-gluon matter occurs, the weak interaction processes for the quarks u, d and s
\begin{equation}
u + s \rightarrow d + u
\, \hspace{1cm} \, and \, \hspace{1cm} \, d + u \rightarrow u + s
\end{equation}
will take place. Since these reactions are relatively
slow, they are not balanced  while the oscillations last and thus, they dissipate
mechanical energy into
heat\cite{wang84}. According to calculations of reference [21], the dissipation
timescale can be estimated by the relation
\begin{equation}
\tau_d \approx 0.01{\left({{150 MeV}\over{m_s}}\right)^4}\left({{M_{\odot}}\over{M_c}}\right) \,\ s
\end{equation}
where $m_s$ is the mass of the s-quark in MeV and M$_c$ is the mass of
the deconfined core in solar masses. This equation is valid for temperatures in
the range $10^8-10^9K$. On the other hand,
the damping timescale by GW emission is
\begin{equation}
\tau_{gw} = 1.8\left({{M_{\odot}}\over{M}}\right)\left({{P^4_{ms}}\over{R^2}}\right)    \,\  s
\end{equation}
where again the stellar mass is in solar units, the radius is in km and the rotation
period P is in milliseconds.

In a first approximation, the fraction of the mechanical energy which will be
dissipated under the form of GWs  is  $f_g = {{1}\over{(1 + \tau_{gw}/\tau_d)}}$.
Notice that the damping timescale by GW emission depends strongly on the
rotation period. Therefore one should expect that slow rotators will dissipate
mostly of the mechanical energy into heat.  In table 4
is given for each star model the expected frequency of the fundamental mode (corrected
for the gravitational redshift) , the critical rotation period (in ms) for having f$_g$ = 0.50, the
GW damping for this critical period and the quality factor of the
oscillation, Q =$\pi\nu_{0}\tau_{gw}$.

Considering now a strongly mangetized neutron star whose $B\sim 10^{18}G$, we
detect an increasing in the gravitational mass of a star composed by pure
hadronic matter. However, the gravitational mass of hybrid stars decrease for
the case of increasing magnetic field and the explanation is quite simple: once
the hybrid star has lower gravitaional mass than the pure hadronic one, due to
the presence of a softer quark matter in its inner shells and, the phase
transition from hadron to quark matter occurs at lower densities with higher
magnetic fields, the highly magnetized hybrid star will have a more important
quark-gluon plasma core and a lower mass. This fact will represent a larger
energy released during the phase transition and gravitational wave emission
which could be detected from larger distances.

\begin{table}[htb]
\begin{center}
\caption{Oscillation parameters for B=0: the damping timescale $\tau_{gw}$ is given for the critical period; maximum
distances for VIRGO (V) and LIGO II (L) are in Mpc, and the LIGO II result for B=$10^{18}G$.}
\vspace{0.5cm}
\begin{tabular}{|lllllll|} \hline 
$\nu_{0}$ &$P_{crit}$ & $\tau_{gw}$&Q&D$_{max}$&D$_{max}$&D$_{max}$\\ 
(kHz)&(ms)&(ms)&-&{(VIRGO)}&{(LIGO II)}&$B=10^{18}G$\\ \hline
1.62&1.64&87.0&442&4.9&10.2&11.5\\ 
1.83&1.25&27.0&155&6.4&13.5&15.28\\ 
2.06&1.13&17.0&110&6.0&12.8&14.48\\
2.32&1.06&11.5&84&5.1&11.1&12.56\\ 
2.72&1.00&8.4&72&3.6&5.7&6.45\\ \hline
\end{tabular}
\end{center}
\end{table}

After filtering the signal, the expected signal-to-noise ratio is

\begin{equation}
{(S/N)^2} = 4\int_0^{\infty}{{\mid \tilde h(\nu)\mid}\over{S_n(\nu)}}d\nu
\end{equation}
where $\tilde h(\nu)$ is the Fourier transform of the signal and
S$_n(\nu)$ is the noise power spectrum of the detector. Performing the required
calculations, the S/N ratio can be written as
\begin{equation}
{(S/N)^2} = {4\over 5}{h_0^2}\left({{\tau_{gw}}\over{S_n(\nu_{gw})}}\right){{Q^2}\over{1 + 4Q^2}}
\end{equation}
In the equation above, the angle average on the beam factors of the detector were already performed.

From eqs.( 4) and (11), once the energy and the S/N ratio are fixed, one can estimate
the maximum distance D$_{max}$ to the source probed by the detector. In the last two columns of
table 4 are given distances D$_{max}$ derived for a signal-to-noise ratio S/N = 2.0
and the sensitivity curve of the laser beam interferometers VIRGO and LIGO II.
In both cases, it was assumed that
neutron stars underwent the transition having a rotation period equal to the critical
value.

We emphasize again that our calculations are based on the
assumption that the deconfinement transition occurs in a dynamical
timescale \cite{haensel82}. In the scenario developed in
ref.\cite{Glendenning}, a mixed quark-hadron phase appears and the
complete deconfinement of the core occurs according to a sequence
of quasi-equilibrium states. The star contracts slowly, decreasing
its inertia moment and increasing the its angular velocity until
the final state be reached in a timescale of the order of $10^5$
yr \cite{Glendenning}. Clearly, in this scenario no gravitational
waves will be emitted and this could be a possibility to
discriminate both evolutionary paths.

\section{Additional Effects}

Additional effects and/or scenarios may improve, or not, the detection
of gravitational waves from compact stars. An important aspect is related 
to the colour superconducting phase. As, in principle, this newer phase would 
turn the equation of state even softer, enlarging the energy release during the
astrophysical phase transition, we decide to analyse such a phase.

The superconducting quark phase is described by the thermodynamical
potential\cite{Lugones}

\begin{equation}
\Omega_{CFL}=\Omega_{Free}-\frac{3}{\pi^2}\Delta^2\mu^2+B \, .
\end{equation}

The resulting equation of state was evaluated in the TOV equations
leading to a smaller radius star, as expected, but also to a less massive
star, even for the gravitational and for the baryonic mass. This
results did not change significantly the properties of gravitational
wave emission as we have expected, however, we still expect a better
understanding about high-density matter properties to better describe
such phase.

The convertion from a neutron star to a strange star, instead of a
hybrid star, is a more catastrophic event and may release much more
energy, being detectable from longer distances. This kind of
transition was studied by Bombaci\cite{Bombaci}, in order to describe
the emission of $\gamma$-ray bursts. Applying our results to describe
the emission of gravitational waves we got the following results which leads to
one of the most promissing gravitational wave generators.

\begin{table}[htb]
\begin{center}
\caption{Properties of a NS$\rightarrow$SS conversion and the maximum
distances for VIRGO (V) and LIGO II (L) are in Mpc}
\vspace{0.5cm}
\begin{tabular}{|lllll|} \hline 
$M_{bar}$ &$M_{SS}$ & $R_{SS}$&D$^{Virgo}_{max}$&D$^{Ligo}_{max}$\\ \hline
1.0749&0.6445&8.40&15.47&34.63\\ 
1.3202&0.7804&8.87&20.01&44.79\\
1.5724&0.9241&9.30&24.27&54.33\\ 
1.8342&1.0727&9.68&28.20&63.13\\
2.1045&1.2210&9.98&32.07&71.79\\ \hline
\end{tabular}
\end{center}
\end{table}

\section{Conclusions\label{c}}

Summarizing, we worked on a nuclear many-body theory with parametrized coupling
constants and the MIT bag model. The phase transition from hadron to quark
matter was described considering screening effects and the equation of state was
determined. We applied the resulting EOS to the TOV equations also considering
intense magnetic field effects. These results are according to previous results
obtained by many authors. As a tool for gravitational wave emission, we
considered the phase transition occuring inside the compact star obtained from
the TOV equations and releasing energy. According to our results, such an event
would be detected as far as 13 Mpc without considering magnetic fields and 15
Mpc for compact star with B=10$^{18}$G. Considering a NS$\rightarrow$SS
transition, we obtained a much larger distance of up to 71 Mpc.

\end{document}